\documentclass[a4paper,12pt]{article} % European paper size
\usepackage{geometry,epsfig,subfigure}
\usepackage{amsmath}
\usepackage{amssymb}
\usepackage{graphics,epsfig,mydef}

\usepackage[square,sort,comma,numbers]{natbib}

\usepackage[usenames]{color}
\usepackage{fancyhdr,url}   % must be loaded for header/footer
\usepackage{setspace}  % must be loaded for \setstretch
\usepackage{multirow}   % must be loaded for \multirow
\usepackage{rotating}    % must be loaded for "sidewaystable" environment
\usepackage{colortbl}     % must be loaded for \columncolor
\usepackage{relsize} % relative sizing of text with smaller/larger
\usepackage{booktabs}
\usepackage{cancel}
\usepackage{enumerate}
\numberwithin{equation}{section}
\usepackage{soul} %for highlights
\usepackage{xcolor} %to cahnge highlight color
\usepackage{pdflscape}
\usepackage{hyperref}
\usepackage[capitalise]{cleveref}
\usepackage{lineno}
\usepackage{graphicx}% Include figure files
\usepackage{etoolbox}
\usepackage{epstopdf}

\definecolor{darkblue}{rgb}{0,0,0.8}
\definecolor{darkgreen}{rgb}{0,0.5,0}

\long\def\symbolfootnote[#1]#2{\begingroup \def\thefootnote{\fnsymbol{footnote}}\footnote[#1]{#2} \endgroup} 

\renewcommand{\cos}[1]{ \text{cos}\hspace{0.0cm}\left( {#1} \right) }

\usepackage{amsthm}

\newcommand{\HRule}{\rule{0.9\linewidth}{0.2mm}}

\usepackage{nomencl}
\makenomenclature

\usepackage{etoolbox}
\renewcommand\nomgroup[1]{%
  \item[
  \ifstrequal{#1}{A}{\textit{Symbols}}{%
  \ifstrequal{#1}{B}{\textit{Greek symbols}}{%
  \ifstrequal{#1}{C}{\textit{Subscripts}}{}}}%
]}
\begin{document}
\renewcommand*{\thepage}{\arabic{page}}

\setstretch{1.3}

\begin{center}
\large
\textbf{On the effect of structural forces on a condensing film profile near a fin-groove corner\\}

\normalsize
\vspace{0.2cm}
Osman Akdag$^{a}$, Yigit Akkus$^{a}$, Zafer Dursunkaya$^{b}\symbolfootnote[1]{e-mail: \texttt{refaz@metu.edu.tr}}\!$\\
\smaller
\vspace{0.2cm}
$^a$ASELSAN Inc., 06200 Yenimahalle, Ankara, Turkey\\
$^b$Department of Mechanical Engineering, Middle East Technical University, 06800 \c Cankaya, Ankara, Turkey\\
\vspace{0.2cm}
\end{center}

\begin{center} \noindent \HRule \\ \end{center}
\vspace{-0.6cm}
\begin{abstract}

\noindent 

Estimation of condenser performance of two-phase passive heat spreaders with grooved wick structures is crucial in the prediction of the overall performance of the heat spreader. Whilst the evaporation problem in micro-grooves has been widely studied, studies focusing on the condensation on fin-groove systems have been scarce.  Condensation on fin-groove systems is actually a multi-scale problem. Thickness of the film near the fin-groove corner can decrease to nanoscale dimensions, which requires the inclusion of nanoscale effects into the modeling. While a few previous studies investigated the effect of dispersion forces, the effect of structural forces has never been considered in the thin film condensation modeling on fin-groove systems. The present study utilizes a disjoining pressure model which considers both dispersion and structural forces. The results reveal that structural forces are able to dominate dispersion forces in certain configurations. Consequently, by intensifying the disjoining pressure, structural forces lead to a sudden change of the film profile (slope break) for subcooling values which are relevant to engineering applications.

\vspace{0.2cm}

\end{abstract}
\vspace{-0.6cm}
\begin{center} \noindent \HRule \\ \end{center}

\pagebreak

%\printnomenclature

\nomenclature[C]{$c$}{capillary}
\nomenclature[C]{$d$}{disjoining}
\nomenclature[C]{$l$}{liquid}
\nomenclature[C]{$lv$}{liquid-vapor}
\nomenclature[C]{$v$}{vapor}
\nomenclature[C]{$w$}{wall}
\nomenclature[C]{$s$}{first derivative with respect to $s$}
\nomenclature[C]{$ss$}{second derivative with respect to $s$}
\nomenclature[C]{$sss$}{third derivative with respect to $s$}

\section{Introduction}
\label{sec:intro}

Thin film condensation is at the center of numerous engineering applications \cite{attinger2014}.  It has been widely studied for typical engineered geometries such as along plates and inside channels \cite{kandlikar2005}. One of the critical geometries, where thin film condensation takes place, is the fin-groove system, which is the representative unit structure for a grooved wall. A considerable amount of two-phase passive heat spreaders such as heat pipes and vapor chambers utilize grooves at their inner walls to convey the liquid between evaporator and condenser sections. Especially, heat pipes with rectangular grooved wicks are an essential part of the thermal management in space applications since they enable the transportation of liquid in long distances due not only to their higher permeability compared to their sintered counterparts \cite{hwang2007,li2016,atay2019} but also their ease of manufacture \cite{faghri1995,wang2005,alijani2019}. Moreover, utilization of embedded micro-grooves is much more feasible compared to the wicks with sintered grains and wire meshes passive in the chip-level electronic cooling applications \cite{peterson1993,harris2010,kundu2015}.

Thermal performance of a two-phase passive heat spreader is basically dictated by its capillary pumping ability together with the phase change efficiency of the evaporator and condenser sections. Despite the abundance of studies focusing on the modeling of evaporation in grooved heat pipes,
 \cite{busse1992,wang_garimella2007,lefevre2008,do2008,lips2010,bertossi2009,do2010,biswal2011,biswal2013,bai2013,ball2013,kou2015,akkus2016,akkus2017}, studies on the condensation modeling remain restricted \cite{lefevre2008,do2008,lips2010,kamotani1976,zhang2001,akdag2019,alipour2019}. In a grooved heat pipe, condensation modeling targets to capture the physics on the fin top and near the fin-groove corner since the condensation intensifies on the fin top because of the small thickness of liquid (micro-scale) compared to the one inside the groove (mili-scale). Moreover, condensed liquid flows into the the groove passing over the fin-groove corner region, where the thickness of liquid film can locally decrease to nanoscale dimensions, which makes the modeling of the condensation further complicated since nanoscale effects can possibly influence the dynamics of the problem. 

Previous attempts towards the modeling of condensation avoided the difficulties associated with multi-scale nature of the problem by adopting several simplifying assumptions. One was to assume a 4\textsuperscript{th} order polynomial for the profile of the liquid-vapor free surface \cite{do2008,kamotani1976}. Another was the assumption of continuous slope of the free surface at the fin top-groove corner \cite{lefevre2008,do2008,lips2010,kamotani1976,alipour2019}. Although these assumptions on the film profile were vastly applied in the literature, their experimental verification has never been made. On the contrary, experimental study of Lips \textit{et al.} \cite{lips2010} reported a \textit{slope break} for the free surface at the fin top-groove corner, which contradicted the commonly accepted simplifying assumptions on the film profile.

Liquid flow in the thin film is basically driven by the capillary pressure gradient along the fin top region. Consequently, the shape of the liquid-gas interface is primarily dictated by the capillary pressure in this region. However, experimental observation of an abrupt change in the film profile near the corner region, where the film thickness can decrease to nanoscale dimensions, suggests the fact that disjoining pressure (or its gradient) can be effective in determining the interface profile. A recent study \cite{akdag2019} incorporated the effect of disjoining pressure into the modeling of thin film condensation and numerically verified that a slope break can be present in the thin film profile, but only under certain conditions such as very small subcooling value, \textit{i.e.} temperature difference between wall and vapor. However, this study \cite{akdag2019} only considered the dispersion force component, which prevented it from drawing a broader conclusion on the effect of disjoining pressure.   

Capillary pressure is associated with the missing bonds of the liquid molecules at the liquid-gas interface. Therefore, it can be considered as a surface phenomenon governed by interaction of liquid and gas molecules. When the thickness of the liquid is sufficiently small, on the other hand, the liquid medium together with its free surface with the gas medium is affected by the solid medium. The overall interaction between liquid molecules and solid atoms can yield a net repulsive force, which resists to the thinning of the film and is generally called as disjoining pressure \cite{israelachvili2015}, or a net attractive force favoring thinning. The interaction between liquid molecules and solid atoms, or disjoning pressure in short, has two major components: i) dispersion forces associated with the long-range van der Waals forces and ii) structural forces associated with the molecular layering of the liquid adjacent to the solid atoms. There are other components of the disjoining pressure arising from the presence of charged particles, surfactants, hydrogen bond etc., however, for the sake of brevity, only dispersion and structural forces are considered, which are always present in the liquid-solid interaction, by selecting a certain fluid-solid couple.       

A standard way to model an interaction between two molecules/atoms is the utilization of Lennard-Jones (LJ) potentials \cite{israelachvili2015}, among which LJ 12-6 form is common: $w(r) \sim [(\sigma / r)^{12}-(\sigma / r)^{6}]$, where $\sigma$ is the finite distance where the repulsive and attractive effects are equal and $r$ is the distance between two molecules. While the first term represents short-range Pauli repulsion due to overlapping electron orbitals, algebraically decaying $1/r^6$ term describes van der Waals forces: i) London dispersion energy between nonpolar molecules, ii) Keesom energy between polar molecules, and iii) Debye energy between a polar and a nonpolar molecule. When the term $1/r^6$ is used to estimate the strength of the van der Waals interaction across a thin medium confined between the planar surfaces of two other media, the resultant interaction force (per area) becomes proportional to $1/\delta^3$, where $\delta$ is the thickness of the intermediate thin medium, which corresponds to the thickness of thin liquid film in the problem of interest. This is the basis for the most common representation of the disjoining pressure utilized in the literature: $p_d=A_d/\delta^3$, where $A_d$ is the dispersion constant. However, this relation reflects only the dispersion effect; therefore, disjoining pressure should be augmented with a component showing the effects of structural forces. It should be noted that the component of the disjoining pressure associated with the Pauli repulsion is negligible and not considered in the current problem since, near the corner region the film is substantially thicker than the adsorbed layer thickness.

Attraction between liquid molecules and the solid wall is maximum near the close proximity of the wall, where liquid molecules are forced to order into quasi-discrete layers due to the wall-force-field effect. As a result, severe density fluctuations of liquid occur near the wall. This phenomenon is known as molecular layering and it is, indeed, experimentally \cite{heslot1989, cheng2001} and computationally \cite{akkus2019molecular,akkus2019atomic} observed. In order to mimic these structural effects in a continuum based modeling effort, in general, an oscillatory and decaying force function can be suggested \cite{israelachvili2015}. There exist many studies \cite{horn1981,jang1995,matsuoka1997,kralchevsky1995,trokhymchuk2001} reporting analytical expressions for the structural component of the disjoninig pressure in the literature. These expressions were reviewed and utilized in the evaporating contact line modeling by Setchi \textit{et al.} \cite{setchi2019}. Their results revealed that the only expression yielding a physically meaningful result was the one developed by Trokhymchuk \textit{et al.} \cite{trokhymchuk2001}. Moreover, this expression estimates higher force values than the others except for very small thickness values ($\delta<1$ nm). Therefore, the current study utilizes the structural force expression developed in \cite{trokhymchuk2001} in order to demonstrate the maximum possible structural effect on the profile of a condensing thin film. 
 
In our previous study \cite{akdag2019}, a framework to comprehensively model the thin film condensation at a fin-groove system was developed to demonstrate the effect of dispersion forces on the film profile for the first time in the literature. The current study takes a further step and aims to investigate the effect of structural forces in addition to the dispersion forces on the thin film condensation at a fin-groove system. The objectives of the present study can be summarized as: i) to develop an understanding on the interplay of relevant molecular forces in the condensing thin film and ii) to reveal their effect on the film profile.

\section{Modeling}
\label{sec:modeling}

\subsection{Problem definition and physical domain}

The thin film condensation of vapor on a grooved planar wall with infinite length and width is of interest in the current study. The condensed liquid is discharged at the bottom of each groove at a flow rate equal to the condensation mass flow rate, which renders the problem steady state. Modeling the condensation problem on a wall with infinite length in axial direction and infinitely many grooves in lateral direction eliminates the effect of liquid flow in the axial direction and ensures symmetrical liquid-vapor interface profile with respect to the central planes of fins and grooves, and thus, reduces the complexity of the problem.

\begin{figure}
\includegraphics[scale=0.8]{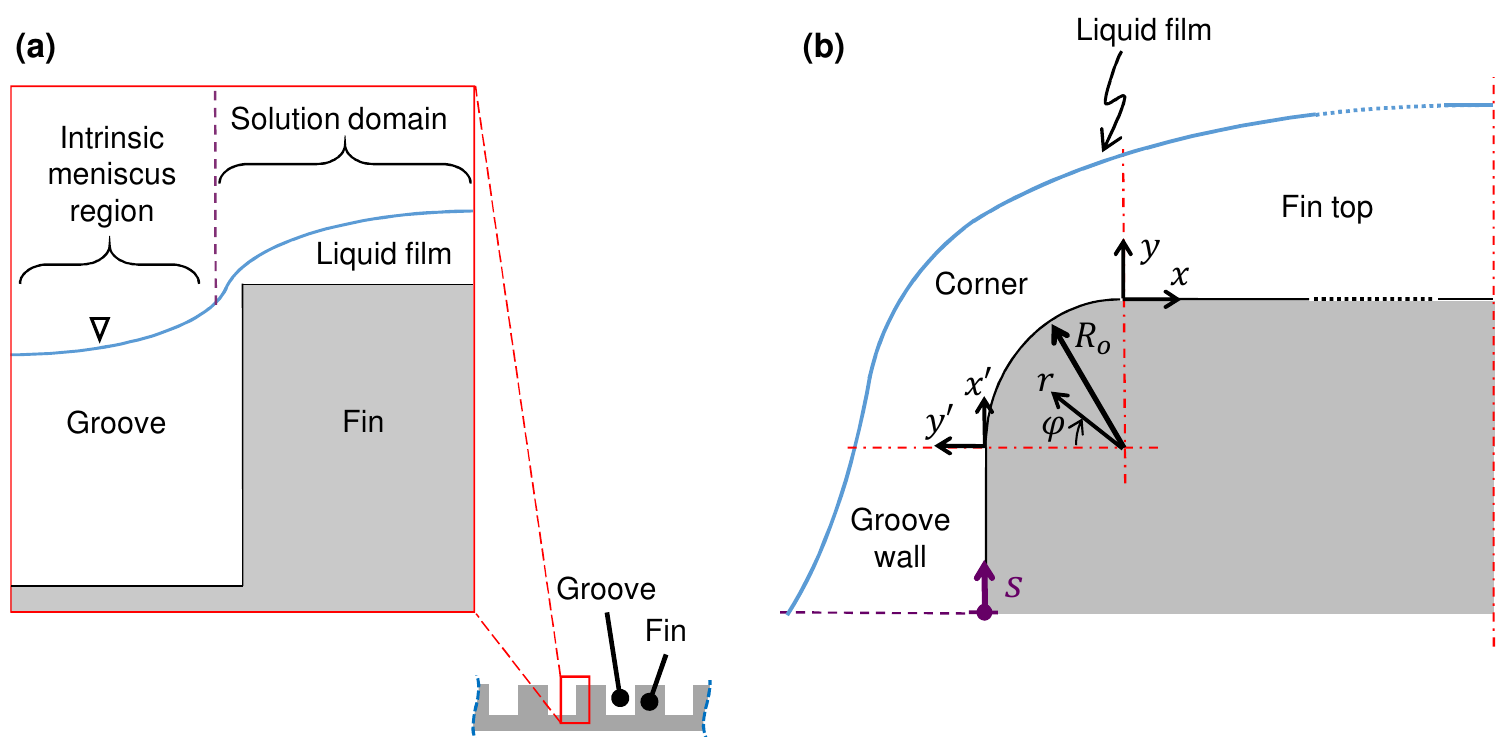}
\centering
\caption{Physical domain. (a) \textit{Problem domain} is defined between the center planes of an adjacent fin and groove pair. (b) \textit{Solution domain} includes the fin top and the close proximity of fin-groove corner extending to the groove.}
\label{problem_domain}
\end{figure}

Neglecting the effects in the axial direction, the condensation problem becomes two-dimensional in the cross-sectional plane of the grooved wall. The region lying in between the center planes---which are the planes of symmetry---of an adjacent fin and groove pair is the problem domain, as shown in Fig.~\ref{problem_domain}a. The condensation predominantly takes place at the fin top surface, where the liquid film thickness is much less compared with the liquid height at the intrinsic meniscus region inside the groove. Therefore, the condensation is modeled on the fin top surface region only. However, the corner region is kept inside the solution domain, since for some cases, the liquid film at the corner region may become excessively thin, which brings up the effect of disjoining pressure and consequently influences the shape of the film forming on the fin top surface, as discussed in~\cite{akdag2019}. Therefore, the solution domain shown in Fig.~\ref{problem_domain}b starts at a point on the groove wall close to the corner, and ends at the central plane of the fin. Solution domain comprises three regions, namely, groove wall, corner and fin top. Corner is assumed as a cylindrical surface. In the formulation, Cartesian~($x$,$y$) coordinates are used for the planar surfaces (groove wall and fin top regions) and polar~($r$,$\varphi$) coordinates are used for the corner region. In the solution algorithm, all the coordinates are transformed to the surface coordinate, $s$. The origins of Cartesian and polar coordinate systems as well as the surface coordinate, $s$, are shown in Fig.~\ref{problem_domain}b.

\nomenclature[A]{\textit{$x$}}{Longitudinal coordinate on fin top region, $\rm m$}
\nomenclature[A]{\textit{$y$}}{Transversal coordinate on fin top region, $\rm m$}
\nomenclature[A]{\textit{$x'$}}{Longitudinal coordinate on groove wall region, $\rm m$}
\nomenclature[A]{\textit{$y'$}}{Transversal coordinate on groove region, $\rm m$}
\nomenclature[A]{\textit{$r$}}{Transversal coordinate on corner region, $\rm m$}
\nomenclature[B]{\textit{$\varphi$}}{Angular coordinate on corner region, $\rm m$}
\nomenclature[A]{\textit{$s$}}{Surface coordinate, $\rm m$}
\nomenclature[A]{\textit{$R_o$}}{Corner radius, $\rm m$}

\subsection{Physical model and governing equations}

In the liquid flow model, the film profile is calculated by solving the mass balance and linear momentum equations throughout the solution domain. The mass balance is based on the relation between the rate of change of mass flow rate per unit depth in $s$-direction, $\dot m'$,  and the mass flux at the free surface due to phase-change (condensation) at the interface, $\dot m_c^{''}$, as follows:  

\begin{equation} 
\frac{{d\dot m'}}{{ds}} =  - \dot m_c^{''}.
\label{mass}
\end{equation}

\nomenclature[A]{\textit{$\dot m'$}}{Mass flow rate per unit depth, $\rm kg \, m^{-1} s^{-1}$}
\nomenclature[A]{\textit{$\dot m_c^{''}$}}{Condensation mass flux, $\rm kg\, m^{-2} s^{-1}$}

The lubrication approximation, which enables the exclusion of inertial and longitudinal diffusive effects~\cite{akdag2019}, reduces the linear momentum equation to:

\begin{subequations}
\begin{equation}
\frac{dp_l}{ds}=\mu\frac{d^2u}{dy^2},
\label{mom-pl}
\end{equation}
\begin{equation}
\frac{dp_l}{ds}=\mu\frac{d^2u}{dr^2},
\label{mom-cyl}
\end{equation}
\end{subequations}

\nomenclature[A]{\textit{$p$}}{Pressure, $\rm Pa$}
\nomenclature[A]{\textit{$u$}}{Velocity in $s$-direction, $\rm m \, s^{-1}$}
\nomenclature[B]{\textit{$\mu$}}{Dynamic viscosity, $\rm kg \, m^{-1} s^{-1}$}

\noindent for planar and cylindrical surfaces, respectively; where, $p_l$ is the liquid pressure, $\mu$ is the dynamic viscosity and $u$ is the velocity in the $s$-direction.  Equation~(\ref{mom-pl}) expresses the conservation of linear momentum at the fin top region and the same equation is used for the groove wall region by replacing $x$ by $x'$  and $y$ by $y'$. The velocity in the $s$-direction can be obtained by integrating Eqs.~(\ref{mom-pl})~and (\ref{mom-cyl}) from the solid wall to the free surface, where no-slip and zero shear boundary conditions are applied, respectively. Then a second integration is performed to obtain the expression for mass flow rate per unit depth:

\begin{equation}
\dot m'=-\frac{1}{3\nu }\frac{dp_l}{ds}\,{\delta}^3,               
\label{m_pl}
\end{equation}
\nomenclature[B]{\textit{$\nu$}}{Kinematic viscosity, $\rm m^2 \rm s^{-1}$}
\nomenclature[B]{\textit{$\nu$}}{Kinematic viscosity, $\rm m^2 \rm s^{-1}$}

\noindent for both planar and cylindrical surfaces. In liquid thin films, the liquid pressure, $p_l$, is a function of interfacial forces, namely, capillary pressure, $p_c$, and disjoining pressure, $p_d$. The augmented Young-Laplace equation defines this relation as:
\begin{equation}
p_v-p_l=p_c+p_d,
\label{ayle}
\end{equation}

\noindent where, $p_v$ is vapor pressure, which is assumed to be constant throughout the domain. The capillary pressure is defined as:

\begin{equation}
p_c=\sigma\kappa ,
\label{pc}
\end{equation}

\nomenclature[B]{\textit{$\sigma$}}{Surface tension, $\rm N \, m^{-1}$}
\nomenclature[B]{\textit{$\kappa$}}{Free surface curvature, $\rm m^{-1}$}

\noindent where, $\sigma$ is the surface tension and $\kappa$ is the curvature of the free surface.

As described previously, the disjoining pressure, $p_d$, arises due to the combination of interfacial forces, among which only the dispersion and structural forces are taken into consideration, since the other components of disjoining pressure are either non-applicable or negligible for the current problem, which considers liquid octane film on a silicon wall. Therefore, the disjoining pressure is defined as:

\begin{equation}
p_d = p_{disp} + p_{str},             
\label{pd}
\end{equation}

\noindent where, $p_{disp}$ is the dispersion component and $p_{str}$ is the structural component. The dispersion component is defined by the well-known power relation:

\begin{equation}
p_{disp}=\frac{A_d}{\delta^3}.             
\label{pdisp}
\end{equation}

\nomenclature[A,1]{\textit{$A_d$}}{Dispersion constant, $\rm J$}

The structural component of disjoining pressure is calculated by following the model presented in~\cite{trokhymchuk2001}:

\begin{equation}
        p_{str}=
        \begin{cases}
            c_1 \cos{c_2 \delta+c_3} \mathrm{e}^{-c_4 \delta } + c_5 \mathrm{e}^{-c_6 \left(\delta-d\right)} & \delta\geq d \\
            -c_7 & \delta\leq d ,  \\
        \end{cases}
\label{eq:n-prime_2}
\end{equation}

\noindent where, $d$ is the molecular diameter and parameters from $c_1$ to $c_6$ can be evaluated based on the bulk fluid properties. It is worth noting that, $c_7$, which is the bulk osmotic pressure of the fluid, is not used since the film thickness, $\delta$, is always greater than the molecular diameter, $d$, which is $0.65 \rm \, nm$ for octane, in the problem solved in the current study. Setchi \textit{et al.}~\cite{setchi2019} evaluated the parameters $c_1$ to $c_6$ for octane at $343 \, \rm K$ as; $c_1=3.95\times10^8 \rm \, Pa$, $c_2=1.08\times10^{10} \rm \, m^{-1}$, $c_3=0.013$, $c_4=6.07 \times 10^8 \rm \, m^{-1}$, $c_5= 2.88 \times 10^{15} \rm \, Pa$, and $c_6=2.21 \times 10^{16} \rm \, m^{-1}$.

The liquid pressure gradient can now be calculated by differentiating the augmented Young-Laplace equation:

\begin{equation}
\frac{dp_l}{ds}=-\frac{d}{ds}\left( p_c+p_d \right).               
\label{dpds}
\end{equation}

The phase-change at the free surface is a function of the temperature difference between the vapor and interface (subcooling) and the pressure difference across the interface (pressure jump) \cite{wayner1971}. Assuming pure conduction inside the liquid film~\cite{moosman1980}, the phase-change mass flux, $\dot m_c^{''}$, is calculated as follows:

\begin{subequations}
\begin{equation}
\dot m_c^{''} = \frac{{a\left( {{T_w} - {T_v}} \right) - b\left( {{p_v} - {p_l}} \right)}}{{1 + 
a\delta h_{lv}/k_l}},
\label{m7}
\end{equation}

\begin{equation}
a= \frac{2c}{2-c} \Bigl(\frac{M}{2\pi R_u T_{lv} } \Bigr)^{1/2}\frac{p_v Mh_{lv}}{ R_u T_v T_{lv}},
\label{m8}
\end{equation}
\begin{equation} 
b= \frac{2c}{2-c} \Bigl(\frac{M}{2\pi R_u T_{lv }} \Bigr)^{1/2}\frac{p_v V_l}{ R_u T_{lv}}, 
\label{m9}
\end{equation}
\end{subequations}

\nomenclature[A]{\textit{$T$}}{Temperature, $\rm K$}
\nomenclature[A,1]{\textit{$h_{lv}$}}{Latent heat, $\rm J \, kg^{-1}$}
\nomenclature[A,1]{\textit{$k_l$}}{Thermal conductivity, $\rm W \, m^{-1} K^{-1}$}
\nomenclature[A,1]{\textit{$c$}}{Accommodation coefficient}
\nomenclature[A]{\textit{$M$}}{Molar mass, $\rm kg \, mol^{-1}$}
\nomenclature[A]{\textit{$R_u$}}{Universal gas constant, $\rm J\, mol^{-1}K^{-1}$}
\nomenclature[A]{\textit{$V_l$}}{Molar volume of liquid, $\rm m^{3} mol^{-1}$}

\noindent where, $T_w$, $T_v$, $T_{lv}$ are the wall, vapor, and liquid-vapor interfacial temperatures, respectively; $h_{lv}$ is the latent heat; $k_l$ is the liquid thermal conductivity; $M$ is the molar mass; $V_l$ is the molar volume of liquid; $R_u$ is the universal gas constant; and $c$ is the accommodation coefficient, which is taken as unity~\cite{do2008,dasgupta1994,du2011,bai2013,kou2015}.

When the expressions for the mass flow rate per unit depth, $\dot m'$, and the condensation mass flux at the free surface, $\dot m_c^{''}$, which are given in Eqs.~(\ref{m_pl}) and~(\ref{m7}), respectively, are substituted into the mass balance equation, Eq.~(\ref{mass}), it takes the following form:
\begin{equation} 
- \frac{1}{{3\nu }}\frac{d}{ds}\Big( \delta^3\frac{dp_l}{ds} \Big) =  - \frac{{a\left( {{T_w} - {T_v}} \right) - b\left( {{p_v} - {p_l}} \right)}}{{1 + a\delta h_{lv}/k_l}},
\label{ode_pl}
\end{equation}
for both planar and cylindrical surfaces. Equation~(\ref{ode_pl}), commonly referred as evolution equation in the literature, is a 4\textsuperscript{th} order ODE of film thickness, solution of which requires four boundary conditions. At the groove side, where $s=0$, the first and second derivatives are calculated based on the radius of curvature of the meniscus, $R_m$, and the edge angle of the liquid-vapor interface inside the groove, $\theta_{g}$, which are known \textit{a priori}. At the end of the solution domain, where $s=L$, the first and third derivatives are zero due to the symmetry condition at this location. The boundary conditions are listed below:

\begin{subequations}
\begin{equation} 
\delta_s = -\tan\theta_g \, , \,\, \delta_{ss}=\frac{\left(1+{\delta_s}^2\right)^{3/2}}{R_m} \,\, \mathrm{at} \,\,  s = 0 \, ,
\label{eq:bc_0}
\end{equation}

\begin{equation} 
\delta_s = 0 \, , \,\, \delta_{sss}=0 \,\, \mathrm{at} \,\, s = L \, .
\label{eq:bc_L}
\end{equation}
\end{subequations}

In the solution algorithm utilized, details of which are given in~\cite{akdag2019}, the film thickness profile is calculated iteratively. The solution starts at the groove side ($s=0$) with the boundary conditions given in Eq.~\ref{eq:bc_0} and the initial guesses of film thickness and mass flow rate per unit depth at $s=0$. Then utilizing two nested secant loops, the film thickness and the mass flow rate per unit depth at $s=0$ rendering the zero slope and zero mass flow rate at the symmetry line---the boundary conditions given in Eq.~(\ref{eq:bc_L})---are sought.

\nomenclature[A]{\textit{$R_m$}}{Radius of curvature of the meniscus, $\rm m$}
\nomenclature[B]{\textit{$\theta_{g}$}}{Edge angle of free surface inside the groove, $^\circ$}

\section{Results and Discussion}

In the simulations, two disjoining pressure models are considered: i) dispersion model, which utilizes only dispersion component ($p_d=p_{disp}$) and ii) structural model, which utilizes both dispersion and structural components ($p_d=p_{disp}+p_{str}$). Fluid properties---evaluated at $343 \, \rm K$ using NIST Chemistry Webbook~\cite{linstrom2001}---and geometrical parameters are given in Table~\ref{props}.

\begin{table}[h]
\caption{Thermophysical properties and geometrical parameters used in the model}

\begin{center}
\begin{tabular}{|l|l|l|}
\hline 
Vapor temperature (K)                     & $ T_v $       & 343 \\
\hline  
Vapor pressure (Pa)                       & $ p_v $       &  15869  \\    
\hline
Density of liquid (kg$\rm \, {m^{-3}}$)      & $ \rho_l $      & 661.38 \\ 
\hline 
Latent heat (J$\rm \, {kg^{-1}}$)         & $ h_{lv} $    & $ 3.398\times 10^{5} $ \\ 
\hline 
Surface tension (N$\rm \, {m^{-1}}$)                      & $\sigma$      & 0.016953 \\ 
\hline
Dynamic viscosity of liquid ($\rm \, Pa \, s$)    & $ \mu _l $       & $ 3.1929\times 10^{-4} $ \\ 
\hline
Thermal conductivity (W$\rm \, {m^{-1}} K^{-1}$)    & $ k_l $       & 0.11136 \\ 
\hline 
Molar mass (kg$\rm \, {mol^{-1}}$)              & $ M $         & 0.11423 \\ 
\hline 
Molar volume of liquid ($ \rm m^{3} \rm {mol^{-1}}$) & $ V_l $       & $1.7271\times10^{-4}$ \\ 
\hline 
Accommodation coefficient                  & $c$ & 1 \\ 
\hline 
Dispersion constant (J)                    &  $ A_d $     & $ 3.18\times 10^{-21} $ \\
\hline  
Molecular diameter (m)                       & $ d $       &  $0.65 \times 10^{-9} $  \\ 
\hline
Radius of meniscus in groove ($\mu$m)                       & $ R_m $        & $ 800 $ \\ 
\hline 
Fin top length ($\mu$m)                       & $ L_{fin} $        & $ 100 $ \\ 
\hline 
\end{tabular} 
 
\end{center}
\label{props}
\end{table}

\nomenclature[B]{\textit{$\rho_l$}}{Density of liquid, $\unit{kg}\unit{m^{-3}}$}
\nomenclature[A,1]{\textit{$L_{fin}$}}{Fin top length, $\unit{m}$}

In our previous work, the same problem was investigated using only dispersion model for subcooling values of $1$, $10^{-1}$, $10^{-2}$, and $10^{-3}\, \rm K$ and the film thickness was found to diminish with decreasing subcooling values due to decreased condensation rates in agreement with previous studies~\cite{khrustalev1994,zhang2001}. The slopes of the film profiles were continuous at the corner region for subcooling values of $1$, $10^{-1}$, and $10^{-2}\, \rm K$, which was in accordance with the commonly utilized continuous slope boundary condition at the corner~\cite{lefevre2008,do2008,lips2010,kamotani1976}. However, a striking result was obtained for near isothermal case ($10^{-3}\, \rm K$ subcooling): a slope break---an abrupt change of film profile slope---near the corner. In near isothermal case, the film was the thinnest and accordingly, the magnitude of disjoining pressure was the highest. Quantitatively, the maximum disjoining pressure in near isothermal case was $5$, $3$, and $2$ orders higher than those in $1$, $10^{-1}$, and $10^{-2}\, \rm K$ subcooling cases, respectively. Therefore, an experimentally identified phenomenon, a slope break~\cite{lips2010}, was numerically confirmed to be present when dispersion forces are effective. The current study, on the other hand, aims to explore the behavior of film profile when both dispersion and structural forces are effective.

\begin{figure}[h]
\includegraphics[scale=0.7]{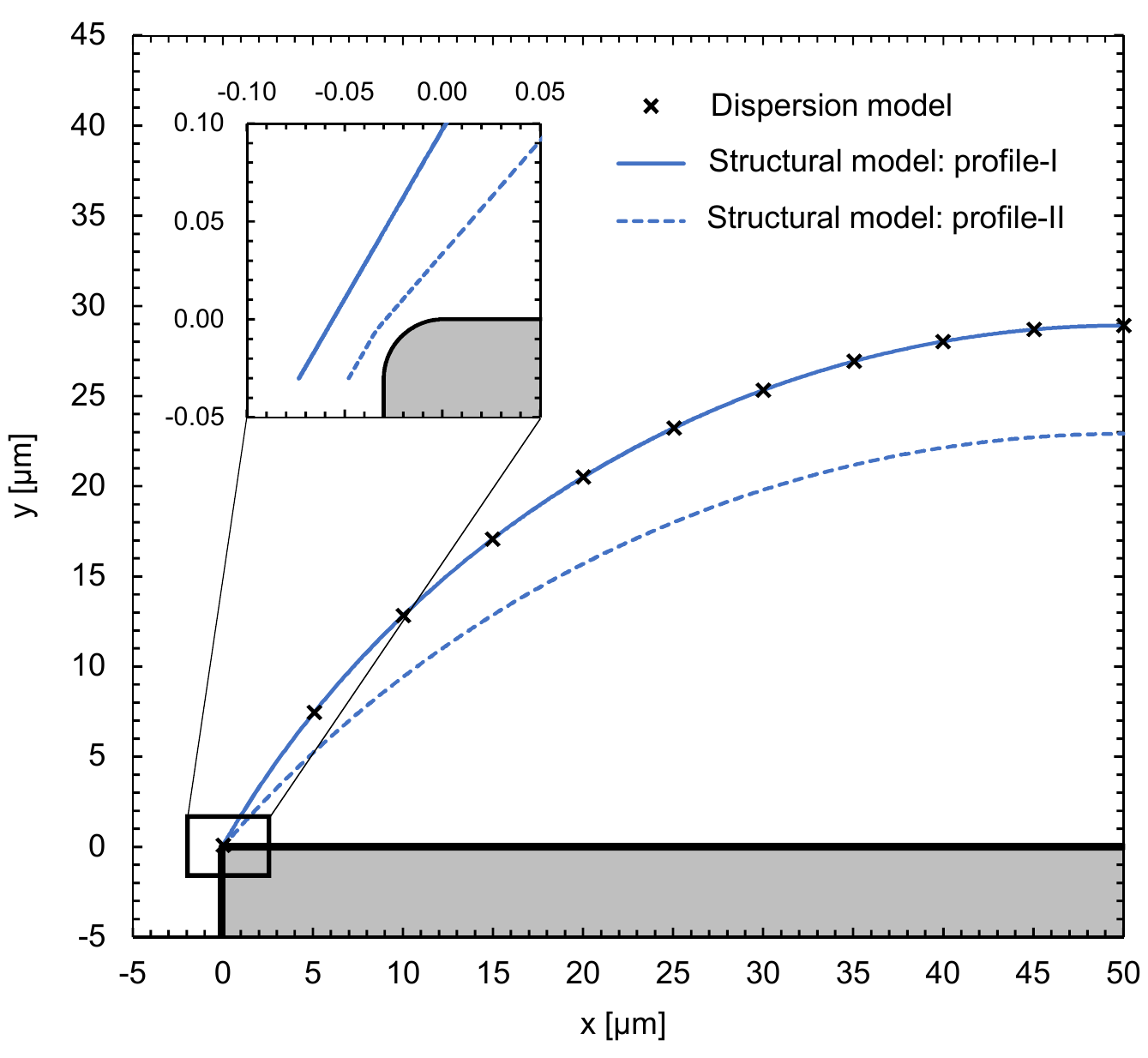}
\centering
\caption{Film profiles obtained for $10^{-1} \, \rm K$ subcooling.}
\label{film_profile}
\end{figure}

\begin{figure}[h]
\includegraphics[scale=0.8]{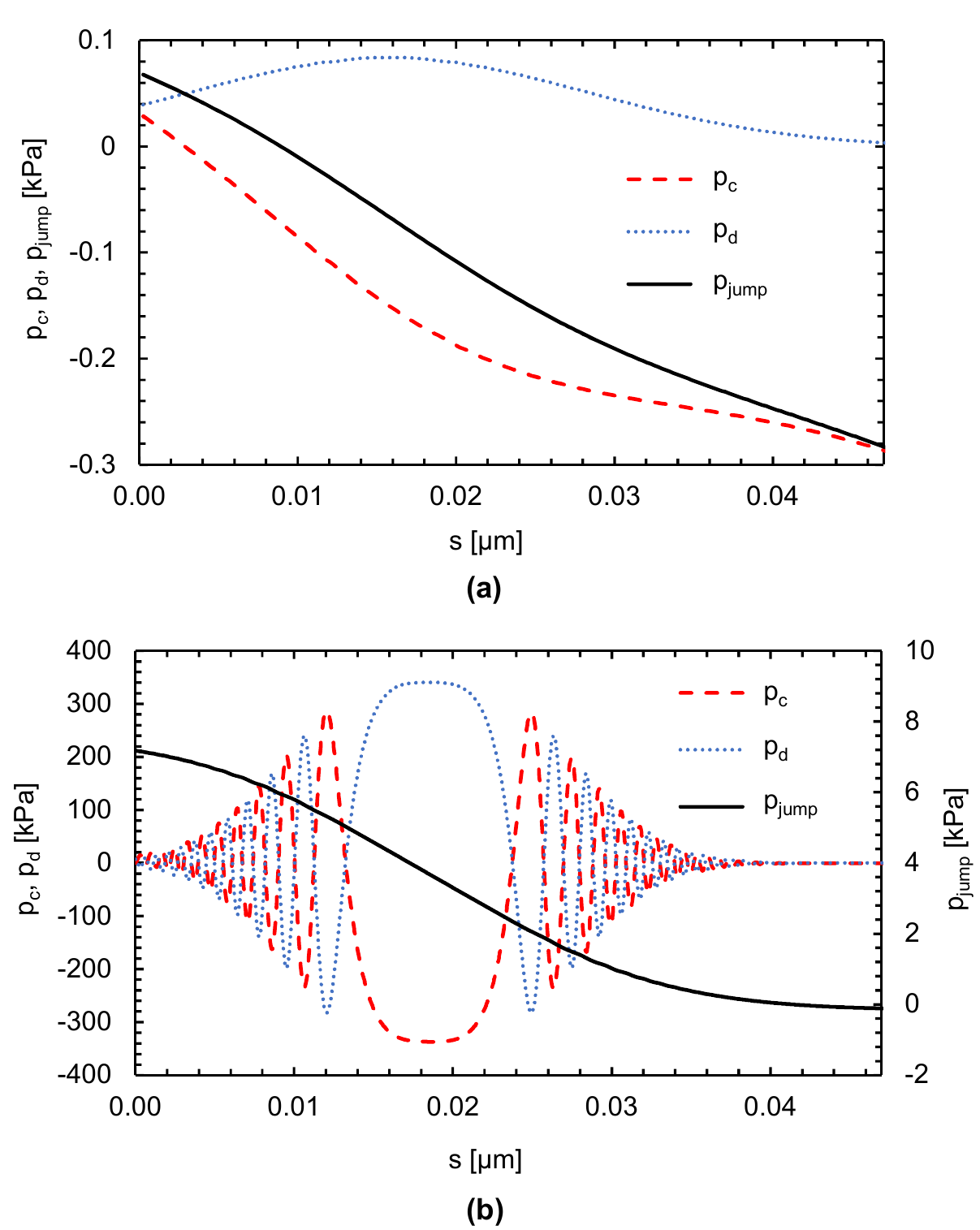}
\centering
\caption{Pressure distributions at the corner region for a) profile-I and b) profile-II. $p_{jump}$ is the summation of capillary ($p_c$) and disjoining ($p_d$) pressures.}
\label{pressure}
\end{figure}

In the first simulation, the condensation problem is solved for $1 \, \rm K$ subcooling using the structural model for disjoining pressure. The algorithm converged to the same result obtained by the dispersion model for this case. However, when the subcooling is decreased to $10^{-1}\, \rm K$, the algorithm converges to two different film profiles, which are given in Fig.~\ref{film_profile}: the first profile (profile-I) being identical to the one obtained with dispersion model; the other one (profile-II)  experiences a strong slope break at the corner. For the same problem with the same boundary conditions, the onset of slope break is identified as $10^{-3}\, \rm K$ when dispersion model is utilized. However, profile-II reveals that slope break can occur even for 2 orders higher subcooling value ($10^{-1} \, \rm K$) as long as the structural forces are included. The slope break leads to the formation of a thinner film at the fin top; and a thinner film implies a higher condensation rate---hence higher heat transfer rate---due to lower thermal resistance. In the current problem, the slope break appearing in profile-II leads to a considerable decrease in the film thickness (from $28.9 \, \rm \mu m$ to $22.9 \, \rm \mu  m$ at the center plane), which results in $13 \%$ higher total condensation rate.

The shape of the free surface is basically determined by the capillary pressure on the fin top surface. However, at the corner region, the film thickness becomes substantially thin giving rise to the disjoining pressure. The pressure difference across the interface is determined by the capillary and disjoining pressures (see Eq.~(\ref{ayle})) and the sudden increase in disjoining pressure at the corner region is compensated by the capillary pressure, which in turn, results in a sudden change in the free surface slope. The pressure distributions at the corner region for profiles-I and -II are presented in Fig.~\ref{pressure}. For profile-I, the film thickness is relatively higher at the corner region, which restricts the effect of disjoining pressure. Therefore, the pressure jump across the interface is very close the capillary pressure, \textit{i.e.} the liquid flow is driven by capillary pressure, as shown in Fig~\ref{pressure}a. However, for profile-II, the magnitude of disjoining pressure at the corner region is high due to the thinner film. Moreover, as seen in Fig~\ref{pressure}b, disjoining pressure has an oscillatory distribution for profile-II suggesting the dominance of structural forces over dispersion forces. Orders of magnitude high and oscillating, disjoining pressure is compensated by an oppositely oscillating capillary pressure as shown in Fig.~\ref{pressure}b. Since the magnitude of the resultant capillary pressure experiences a sudden rise at the corner region, free surface curvature changes abruptly leading to a slope break in the film profile. Despite the oscillations in capillary and disjoining pressures, their sum yields a non-oscillating pressure jump (see Fig.~\ref{pressure}b) across the interface, which implies a smooth liquid pressure distribution, sustaining the liquid flow along the corner region.

Although profile-II has a smooth concave shape in Fig.~\ref{film_profile}, oscillating capillary pressure implies a continuously changing curvature direction at the corner region, which is an indicator of nanoscale wiggles at the free surface shape. A wiggling free surface was also reported by Setchi \textit{et al.}~\cite{setchi2019} at the adsorbed layer of an evaporating thin film, where the same structural force model \cite{trokhymchuk2001} was utilized.

The oscillatory nature of the structural forces renders the solution sensitive to small changes in film thickness at the corner region. Therefore, obtaining a converged solution becomes more difficult as the effect of structural forces intensifies, and the solution becomes chaotic. Another consequence of oscillating structural force is the presence of a multiplicity of results for identical cases: two different film profiles, profiles-I and profile-II, are obtained for the same boundary conditions of the current problem due to minute perturbations in the iterative process. Yet these two film profiles may not be unique, since perturbations can possibly yield entirely different solutions for the film profiles. Regardless of the existence of other possible solutions, as one of the two converged profiles in this study, profile-II manifests the prominent effect of structural forces in the current condensation problem \textit{via} increasing the magnitude of subcooling on the onset of slope break by 2 orders of magnitude. This result reveals that the slope break of film profile can be observed not only in near isothermal cases but also in the cases with relatively higher subcooling values which are relevant to engineering applications.

It can be concluded that the modeling of thin film condensation phenomena when the order of film thickness approaches nanometers, should account not only for the dispersion but also the structural component, the omission of which results in entirely different solutions.  Care should, however, be exercised as the differences in solution approach may lead to a multiplicity of results, which lack sufficient experimental validation.

\addcontentsline{toc}{section}{References}

\bibliographystyle{unsrt}

\bibliography{references}

\section*{Declarations of interest}
\addcontentsline{toc}{section}{Competing_interests}
None.
%%%%%%%%%%%%%%%%%%%%%%%%%%%%%%%%%%%%%%%%%%%%%%%%%%%%%%%%%%%%%%%%%%%%%%

\end{document}